\documentclass[12pt]{article}
\usepackage{epsfig,graphicx}
\usepackage{fpcp03}
\usepackage{multirow}

\newcommand{\BABARPubYear}    {03}
\newcommand{\BABARConfNumber} {65}
\newcommand{\SLACPubNumber} {10190}

\def\beq{\begin{equation}}
\def\eeq#1{\label{#1}\end{equation}}
\def\eeqn{\end{equation}}

\def\beqa{\begin{eqnarray}}
\def\eeqa#1{\label{#1}\end{eqnarray}}
\def\eeqan{\end{eqnarray}}

\let\bar=\overbar

\def\etal{{\it et al.}}

\def\Dslash{\not{\hbox{\kern-4pt $D$}}}
\def\dslash{\not{\hbox{\kern-2pt $\del$}}}

\def\msb{{\bar{\ssstyle M \kern -1pt S}}}

\def\BB0bar{B^0 {\overline B}^0}
\def\BB0dbar{B_d^0 {\overline B}_d^0}
\def\BB0sbar{B_s^0 {\overline B}_s^0}

\def\rar{\rightarrow}

\RequirePackage{xspace}

\usepackage{relsize}
\def\babar{\mbox{\slshape B\kern-0.1em{\smaller A}\kern-0.1em
    B\kern-0.1em{\smaller A\kern-0.2em R}}}

\def\piz   {\ensuremath{\pi^0}\xspace}

\def\pip   {\ensuremath{\pi^+}\xspace}
\def\pim   {\ensuremath{\pi^-}\xspace}

\def\Kbar  {\kern 0.2em\overline{\kern -0.2em K}{}\xspace}

\def\Kz    {\ensuremath{K^0}\xspace}
\def\Kzb   {\ensuremath{\Kbar^0}\xspace}
\def\KzKzb {\ensuremath{\Kz \kern -0.16em \Kzb}\xspace}
\def\Kp    {\ensuremath{K^+}\xspace}
\def\Km    {\ensuremath{K^-}\xspace}

\def\KpKm  {\ensuremath{\Kp \kern -0.16em \Km}\xspace}
\def\KS    {\ensuremath{K^0_{\scriptscriptstyle S}}\xspace}

\def\Dbar    {\kern 0.2em\overline{\kern -0.2em D}{}\xspace}

\def\Dz      {\ensuremath{D^0}\xspace}
\def\Dzb     {\ensuremath{\Dbar^0}\xspace}
\def\DzDzb   {\ensuremath{\Dz {\kern -0.16em \Dzb}}\xspace}
\def\Dp      {\ensuremath{D^+}\xspace}
\def\Dm      {\ensuremath{D^-}\xspace}

\def\DpDm    {\ensuremath{\Dp {\kern -0.16em \Dm}}\xspace}

\def\Dstarp  {\ensuremath{D^{*+}}\xspace}

\def\Bbar    {\kern 0.18em\overline{\kern -0.18em B}{}\xspace}

\def\BB      {\ensuremath{B\Bbar}\xspace} 
\def\Bz      {\ensuremath{B^0}\xspace}
\def\Bzb     {\ensuremath{\Bbar^0}\xspace}
\def\BzBzb   {\ensuremath{\Bz {\kern -0.16em \Bzb}}\xspace}
\def\Bu      {\ensuremath{B^+}\xspace}
\def\Bub     {\ensuremath{B^-}\xspace}

\def\BpBm    {\ensuremath{\Bu {\kern -0.16em \Bub}}\xspace}

\mathchardef\Upsilon="7107
\def\Y#1S{\ensuremath{\Upsilon{(#1S)}}\xspace}

\mathchardef\Deltares="7101
\mathchardef\Xi="7104
\mathchardef\Lambda="7103
\mathchardef\Sigma="7106
\mathchardef\Omega="710A

\def\Deltabar{\kern 0.25em\overline{\kern -0.25em \Deltares}{}\xspace}
\def\Lbar{\kern 0.2em\overline{\kern -0.2em\Lambda\kern 0.05em}\kern-0.05em{}\xspace}
\def\Sigbar{\kern 0.2em\overline{\kern -0.2em \Sigma}{}\xspace}
\def\Xibar{\kern 0.2em\overline{\kern -0.2em \Xi}{}\xspace}
\def\Obar{\kern 0.2em\overline{\kern -0.2em \Omega}{}\xspace}
\def\Nbar{\kern 0.2em\overline{\kern -0.2em N}{}\xspace}
\def\Xb{\kern 0.2em\overline{\kern -0.2em X}{}\xspace}

\def\BR         {{\ensuremath{\cal B}\xspace}}

\newcommand{\tev}{\ensuremath{\mathrm{\,Te\kern -0.1em V}}\xspace}
\newcommand{\gev}{\ensuremath{\mathrm{\,Ge\kern -0.1em V}}\xspace}
\newcommand{\mev}{\ensuremath{\mathrm{\,Me\kern -0.1em V}}\xspace}
\newcommand{\kev}{\ensuremath{\mathrm{\,ke\kern -0.1em V}}\xspace}
\newcommand{\ev}{\ensuremath{\mathrm{\,e\kern -0.1em V}}\xspace}
\newcommand{\gevc}{\ensuremath{{\mathrm{\,Ge\kern -0.1em V\!/}c}}\xspace}
\newcommand{\mevc}{\ensuremath{{\mathrm{\,Me\kern -0.1em V\!/}c}}\xspace}
\newcommand{\gevcc}{\ensuremath{{\mathrm{\,Ge\kern -0.1em V\!/}c^2}}\xspace}
\newcommand{\mevcc}{\ensuremath{{\mathrm{\,Me\kern -0.1em V\!/}c^2}}\xspace}

\def\invfb   {\ensuremath{\mbox{\,fb}^{-1}}\xspace}

\def\mus  {\ensuremath{\rm \,\mus}\xspace}

\def\mus        {\ensuremath{\,\mu{\rm s}}\xspace}    

\def\to                 {\ensuremath{\rightarrow}\xspace}

\def\pep2{PEP-II}

\newcommand{\chisq}{\ensuremath{\chi^2}\xspace}

\def\gsim{{~\raise.15em\hbox{$>$}\kern-.85em
          \lower.35em\hbox{$\sim$}~}\xspace}
\def\lsim{{~\raise.15em\hbox{$<$}\kern-.85em
          \lower.35em\hbox{$\sim$}~}\xspace}

\def\CP                {\ensuremath{C\!P}\xspace}

\xspace

\newcommand{\jprlBase}       {Phys.\ Rev.\ Lett.\xspace}
\newcommand{\jprBase}        {Phys.\ Rev.\xspace}

\newcommand{\jprl}      [1]  {\jprlBase\ {\bf #1}}
\newcommand{\jprd}      [1]  {\jprBase\ D~{\bf #1}}

\def\jetset74   {\mbox{\tt Jetset \hspace{-0.5em}7.\hspace{-0.2em}4}\xspace}

\def\Ampbar{\kern 0.18em\overline{\kern -0.18em A}}

\begin{document}

\begin{flushright}
\babar-PROC-\BABARPubYear/\BABARConfNumber \\
SLAC-PUB-\SLACPubNumber \\
\end{flushright}


\Title{ Branching fractions and direct CP violation\\
in B \boldmath $\rightarrow$ \unboldmath PP(PV) decays }
\bigskip


%
\label{BonaStart}

%
\author{ Marcella Bona\index{Bona, M.} }

%
\address{Universit\`a di Torino and INFN, Sezione di Torino\\
via Pietro Giuria 1 \\
10125 Torino, Italy \\
}

\makeauthor\abstracts{
I present the results of searches for B meson decays into two
charmless mesons. I take into account final states made up of two
pseudo-scalar (PP) or one pseudo-scalar and one vector (PV).
The measurements use the data samples collected and analysed by the 
B-factory experiments at the $\Upsilon(4S)$ resonance energy:
\babar, Belle and CLEO.
}

\section{Introduction}

Charmless hadronic final states play an important role in the study of \CP-violation.
In the Standard Model, all \CP-violating phenomena are a consequence of a single
complex phase in the Cabibbo-Kobayashi-Maskawa (CKM) quark-mixing matrix~\cite{ckm}.

Charmless decays of $B$ mesons may proceed by $b\to u$, $b\to s$,
or $b\to d$ transitions. The latter two mechanisms require flavor
changing neutral currents which are not present at tree level in the
Standard Model, and therefore must occur through suppressed processes
such as the penguin mechanism. Such processes involve loops, which can
get contributions from physics beyond the Standard Model.
Even in the absence of such new physics, interference among competing
amplitudes for a given decay mode can be exploited to measure CKM
phases.

These charmless $B$ decays in which CKM favored amplitudes are suppressed
or forbidden are rare decays and they typically have branching
fractions (\BR) of less than $10^{-4}$.
Even if the branching fraction for these modes is expected to be so low,
the present data samples available to \babar\ and Belle Collaborations
together with the new analysis effort by CLEO allow for measurements or
stringent limits on many such modes.

In these years, the \babar\ and Belle collaborations have reached solid
results~\cite{sin2beta} on measurements of \CP-violating asymmetries 
in $B$ decays into final states containing charmonium, leading to strict
constraints on the angle $\beta$ of the CKM Unitarity Triangle.

In the Standard Model picture, the angle $\alpha$ can be related to
time-dependent \CP-violating asymmetries in the analysis of the decay
$\Bz\to\pip\pim$ as well as in the case of three body $\pi^+\pi^-\pi^0$ decays
in which such measurement would exploit interference between the
$B^0 \rightarrow \rho^{\pm}\pi^{\mp}$ modes and the color-suppressed
$B^0 \rightarrow \rho^0\pi^0$.

Ratios of branching fractions for various $\pi\pi$ and $K\pi$ decay modes are
in principle sensitive to the angle $\gamma$~\cite{beneke}, even if the measured
branching fractions of all the $K\pi$ modes show a clear path which
seems to suggest the dominance of the penguin amplitude over the others~\cite{ciuchini}.
As a matter of fact the $K\pi$ modes (together with the relative PV modes like
$K^*\pi$, $\rho K$, $\omega K$) are sensitive to $\gamma$ through the
Cabibbo-suppressed amplitude term which is proportional to $V_{ub}^*V_{us}$
but an interference between the latter and the Cabibbo-allowed amplitude term is
needed in order to extract $\gamma$\footnote{For an example on this amplitude formalism
see~\cite{buras-silvestrini}}.

The extraction of $\alpha$ from measurements of the time-dependent asymmetry in
$\Bz\to\pip\pim$ is complicated by the interference of tree and penguin amplitudes
with different weak phases. The $B \to KK$ decays are characterized by similar
penguin processes and, hence, can be used to isolate the tree and penguin
contributions to $\Bz\to\pip\pim$.
For example $K^0\bar K^0$ is a pure $b \rightarrow d$ penguin.
The fact that no tree amplitude is possible, strongly reduces the expected
branching fraction, but also allows to estimate the size of penguin contributions in
$B \rightarrow \pi \pi$, once the ${\BR}(B \to K^0\bar K^0)$ has been
measured and SU(3) breaking effects have been taken into account.
$K^{\pm} \bar K^0$ has the same penguin as $K^0\bar K^0$ plus an annihilation
contribution. The comparison of the two branching fractions can be used
to understand which is the real contribution coming from annihilation terms.
Moreover, the entire $K^0\bar K^0$ amplitude is equal to the penguin contribution
to the $\piz\piz$ channel, assuming SU(3). So, in the scenario of a large branching
fraction for $\piz\piz$, the measurement of the ${\BR}(K^0\bar K^0)$ can help to
disentangle penguin contribution from other subleading terms in $\piz\piz$ amplitude.
These other terms can be deduced in a similar way from $K^+K^-$ decay
which is a pure annihilation mode~\cite{buras-silvestrini}.

Also the decays of $B$-mesons to pseudo-scalar and vector particles provide
opportunities for investigating the phenomenon of $CP$ violation. Of particular
interest are the quasi-two-body $B \to \rho \pi$ decays to three-pion final
states. These can be used to measure the angles $\alpha$ and $\gamma$.

$\alpha$ can be determined from a full Dalitz plot analysis of the modes
$B^0 \to \rho^\pm \pi^\mp$ and $B^0 \to \rho^0 \pi^0$~\cite{pipipi-phi2}.
The data sets available are still too small for this type of analysis.
For the time being though, the \babar\ experiment also performed a
simultaneous measurement of branching fractions and \CP-violating
asymmetries in the decays $B^0\rar\rho^{\pm}\pi^{\mp}$ 
and $B^0\rar\rho^{-} K^{+}$ (and their charge conjugate)
using a time-dependent maximum likelihood analysis. 
Following a quasi-two-body approach~\cite{BaBarPhysBook},
the \babar\ technique is restricting the analysis to the two regions of the 
$\pi^{\mp}\pi^0 h^{\pm}$ Dalitz plot ($h=\pi$ or $K$) that are dominated by
either~$\rho^{+}h^{-}$ or~$\rho^{-}h^{+}$. 
From this kind of analysis, the time-integrated and flavor-integrated charge
asymmetries $A_{CP}^{\rho\pi}$ and $A_{CP}^{\rho K}$
that measure direct \CP\ violation can be extracted together with
the quantities $S_{\rho\pi}$ and $C_{\rho\pi}$ for the $\rho\pi$ mode.

Since, like in the two-body case, the extraction of $\alpha$ from $\rho^\pm\pi^\mp$
is complicated by the interference of decay amplitudes with different weak and strong phases,
another strategy is to perform an SU(2) analysis using all $\rho\pi$ final states~\cite{nir}.
Assuming isospin symmetry, the angle $\alpha$ can be determined free of hadronic
uncertainties from a pentagon relation formed in a complex plane by the five decay
amplitudes $B^0\rar\rho^+\pi^-$, $B^0\rar\rho^-\pi^+$, $B^0\rar\rho^0\pi^0$,
$B^+\rar\rho^+\pi^0$ and $B^+\rar\rho^0\pi^+$.
These amplitudes can be determined from measurements of the corresponding decay rates and
\CP-asymmetries.
$\gamma$ can be extracted from the interference of
$B^+ \to \chi_{\mbox{\tiny c0}} \pi^+$ and $B^+ \to \rho^0 \pi^+$ in 
$B^+ \to \pi^+ \pi^- \pi^+$ decays~\cite{pipipi-phi3}.

Also measurements of the charge asymmetries in the decay rates are reported
according to the definition:
\[
{\cal A}_{CP}=\frac{
\Gamma\left(\Bbar\to\bar{f}\right)-\Gamma\left(B\to f\right)}
{\Gamma\left(\Bbar\to\bar{f}\right)+\Gamma\left(B\to f\right)}
\,.
\]

Here and throughout this paper charge conjugate modes are implied. We
also make use of the notation $h^\pm$ to represent a charged hadron that may be
either a kaon or pion.

\section{Analyses}

The measurements presented in this paper are mainly based on the following data samples.
The analyses of \babar\ are based on $(87.9 \pm 1.0) \times 10^{6}$ $B\bar B$ pairs collected
between $1999$ and $2002$ with the \babar\ detector at the PEP-II asymmetric-energy $B$
Factory at SLAC. This corresponds to an integrated luminosity of approximately 
$81 \invfb$ at the $\Upsilon(4S)$ resonance.
The results from Belle are based on $78 \invfb$ data sample which corresponds to
$85 \times 10^{6}$ $B\bar B$ pairs. The Belle detector\cite{nimbelle} is a large-solid-angle
general purpose spectrometer at the KEKB asymmetric-energy $e^+e^-$ storage ring.
The current CLEO data set is made up from the full CLEO II and CLEO III data samples totaling
$15.3 \invfb$. CLEO\cite{nimcleo} is a general purpose solenoidal magnet detector operated at the
Cornell Electron Storage Ring (CESR), a symmetric-energy storage ring tuned to provide center of
mass energies near the $\Upsilon(4S)$ resonance.

Hadronic events are selected based on track multiplicity and event topology.
Tracks are identified as pions or kaons using the various Particle Identification
(PID) techniques described below.
Candidate $\KS$ mesons are reconstructed from pairs of oppositely charged tracks that
form a well-measured vertex and have an invariant mass within a defined window around
the nominal \KS\ mass~\cite{PDG}. The candidate must have a displaced vertex and flight 
direction consistent with a $\KS$ originating from the interaction point.
Candidate \piz\ mesons are formed from pairs of photons with an invariant mass within
a defined window around the nominal \piz\ mass. The \piz\ candidates are then
kinematically fitted with their mass constrained to the nominal \piz\ mass~\cite{PDG}.

\begin{figure}[t]
\begin{center}
\begin{minipage}[tbh]{16.0cm}
\epsfig{file=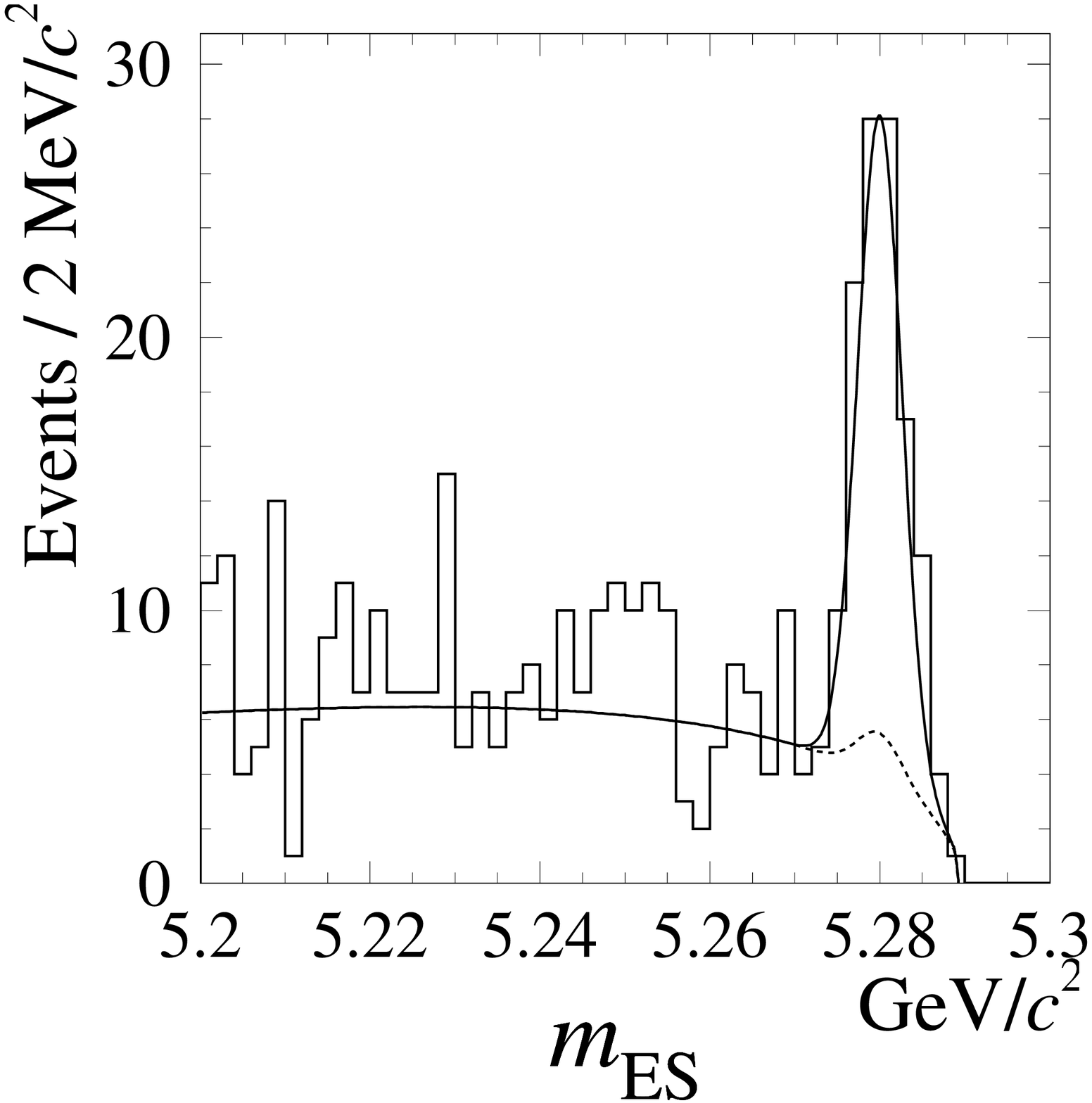,height=5cm}
\epsfig{file=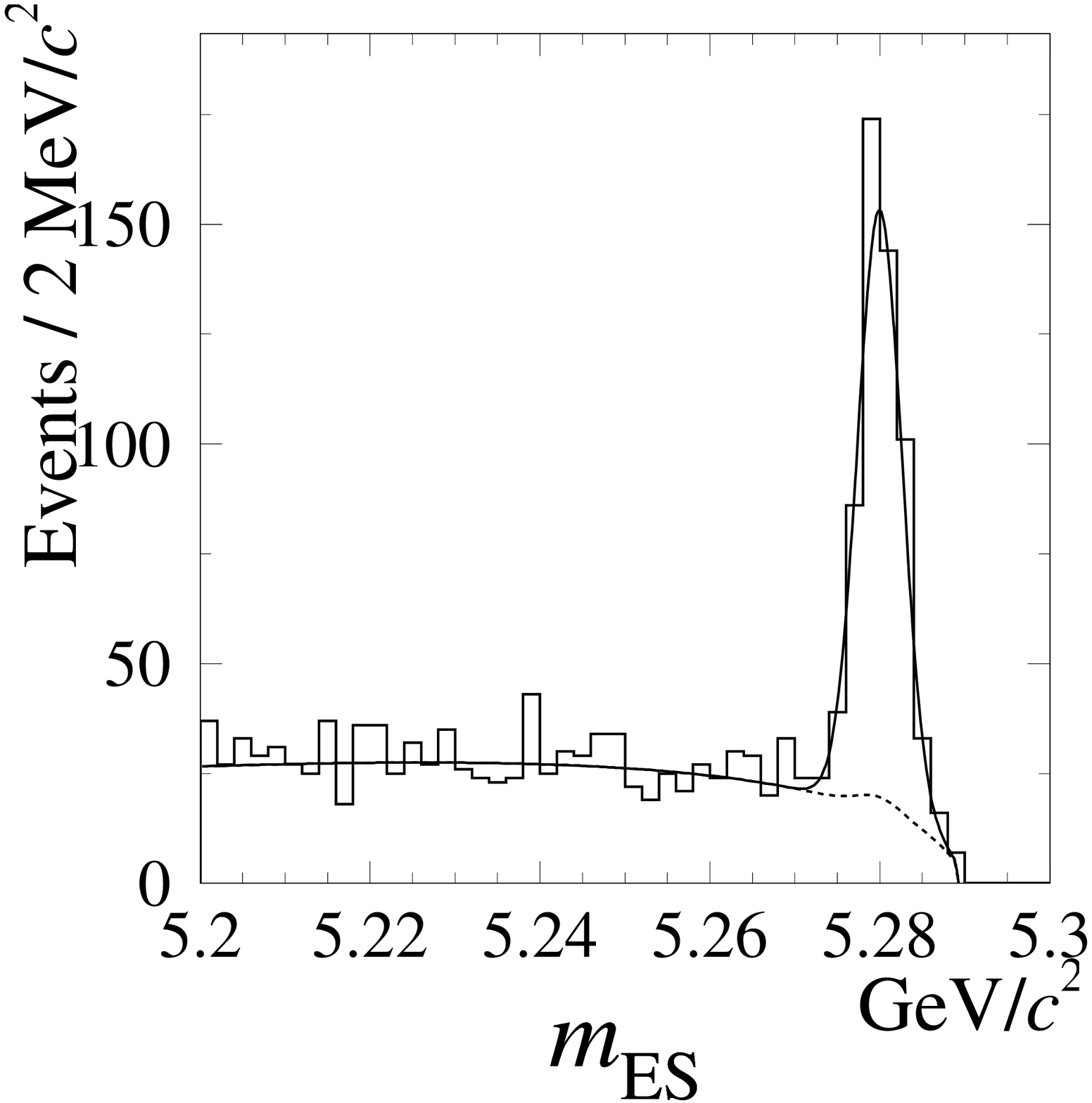,height=5cm}
\epsfig{file=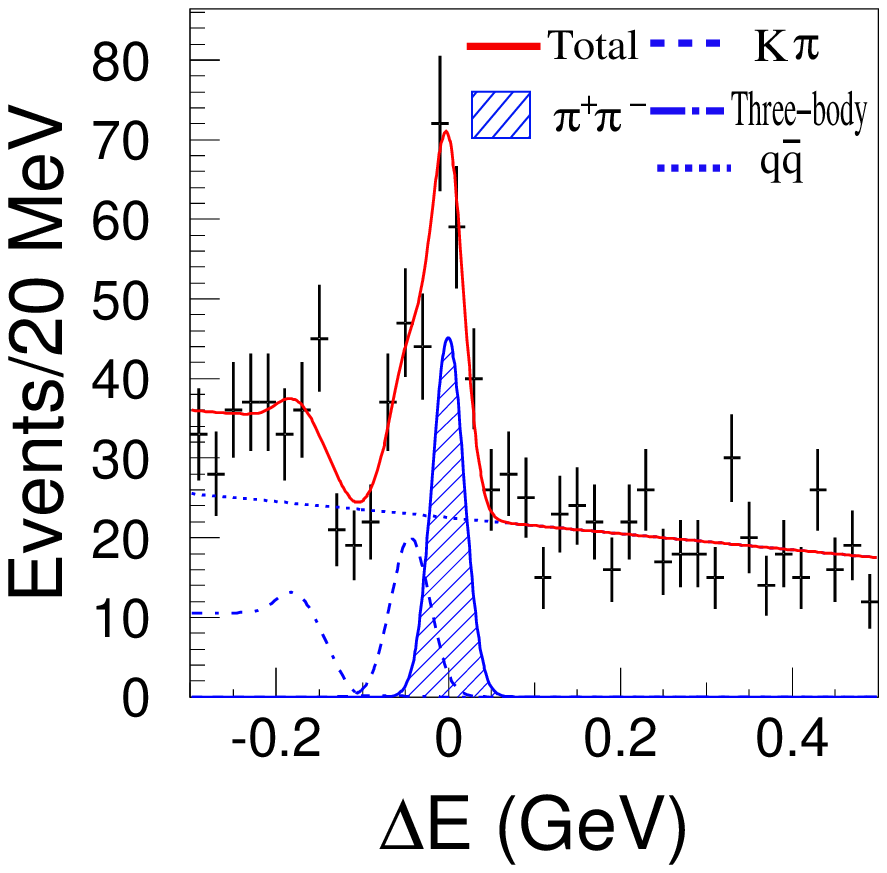,height=5cm}
\end{minipage}
\begin{minipage}[tbh]{16.0cm}
\hspace{0.6cm}
\epsfig{file=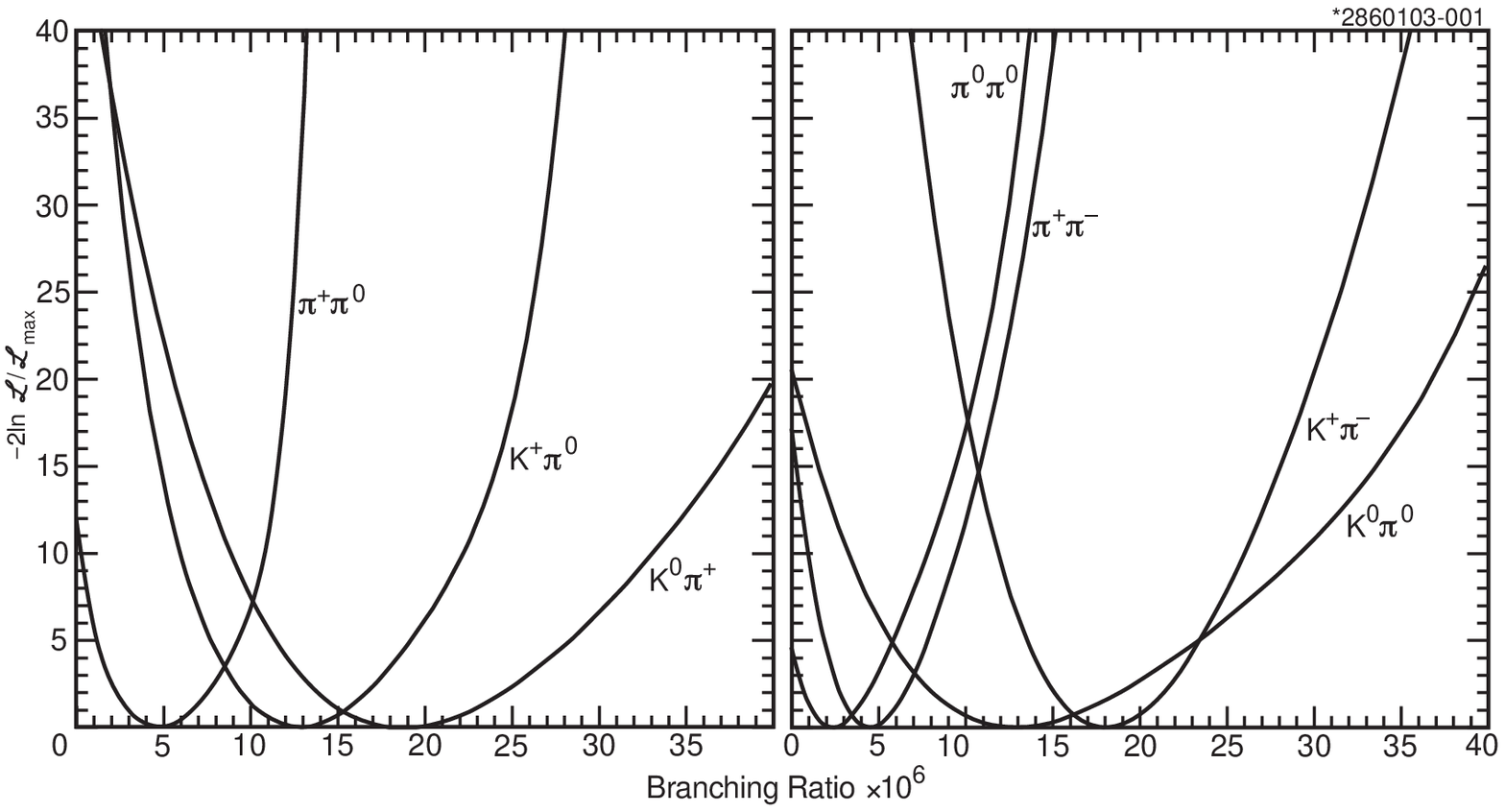,height=5.cm}
\hspace{0.15cm}
\epsfig{file=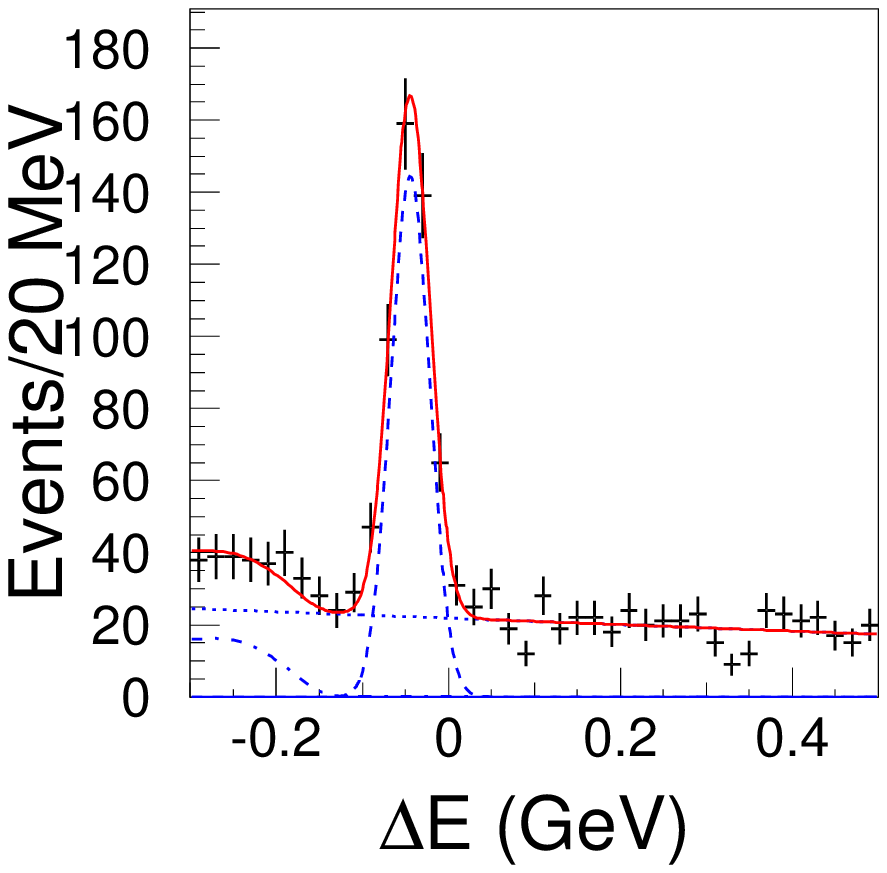,height=5cm}
\end{minipage}
\caption{Examples of plots from the various two-body analyses and techniques.
See text in Sect.~\ref{results} for details.}
\label{fig:twobody}
\end{center}
\end{figure}

Reconstructed $B$ candidates are identified using two kinematic variables:
the beam-energy substituted mass (or beam-energy constrained mass)
$m_{ES} = \sqrt{(E_{\rm beam}^{\rm CM})^2 -(p_{B}^{\rm CM})^2}$ and the energy
difference $\Delta E = E_{B}^{\rm CM}-E_{\rm beam}^{\rm CM}$, where $E_{\rm beam}^{\rm CM}$
is the beam energy, $p_{B}^{\rm CM}$ and $E_{B}^{\rm CM}$ are the momentum and
the energy of the reconstructed $B$ meson in the center-of-mass system (CM).

In these charmless modes, the largest source of background comes from random
combinations of tracks and neutrals produced in the $e^+e^- \rightarrow q\bar q$
continuum (where $q=u$, $d$, $s$ or $c$). These backgrounds are suppressed using
the event topology. In the CM frame this background typically exhibits
a two-jet structure that can produce two high momentum, nearly
back-to-back particles, in contrast to the spherically symmetric
nature of the low momentum $\Upsilon(4S) \to b\bar b$ events.
This topology difference is exploited by making use of event-shape quantities.
Various techniques and variables are used.

In the \babar\ analyses, a preliminary requirement is applied on the angle
$\theta_{sph}$~\cite{spheric} between the sphericity axes, evaluated in the CM frame,
of the $B$ candidate and the remaining tracks and photons in the event.
The distribution of the absolute value of $\cos\theta_{sph}$ is strongly peaked near $1$ 
for continuum events and is approximately uniform for \BB\ events.

Another similarly powerful variable is the flight direction of the
$B$ candidate given by $\cos\theta_B = \hat{\bf p}\cdot \hat{\bf z}$
where ${\bf p}$ is the vector sum of the daughter momenta and $\bf z$ is
the beam axis. Since the vector $\Upsilon(4S)$ is produced in $e^+e^-$ annihilation
it has a polarization $J_z=\pm 1$, and the subsequent flight direction of
the pseudo-scalar $B$ mesons is distributed as
$|Y^{\pm 1}_1(\theta,\phi)|^2\sim\sin^2\theta = 1-\cos^2\theta$. 
Background events are flat in this variable.

Another quantity largely used in the charmless analyses is a Fisher discriminant
${\cal F}$ which consists of an optimized linear combination of 
variables that distinguish signal from background~\cite{fisher}.

\babar\ analyses use a two-variable Fisher discriminant with $\sum_i p_i$
and $\sum_i p_i|\cos \theta_i|^2$ where $p_i$ is the momentum and
$\theta_i$ is the angle with respect to the thrust axis of the $B$
candidate, both in the CM frame, for all tracks and neutral clusters
not used to reconstruct the $B$ meson.

Belle's technique consists of forming signal and background
likelihood functions, ${\cal L}_S$ and ${\cal L}_{BG}$, 
from two variables. One is a Fisher
discriminant determined from six modified Fox-Wolfram
moments~\cite{foxbelle} and the other is $\cos\theta_B$.
Requirements are imposed on the likelihood ratio
$LR = {\cal L}_S/({\cal L}_S+{\cal L}_{BG})$ for candidate events.

CLEO builds its Fisher discriminant as a linear combination of
fourteen variables including the direction of the thrust
axis of the candidate with respect to the beam axis,
$\cos\theta_{thr}$, and the nine conical bins of a ``Virtual Calorimeter''.
They are the scalar sum of the momenta of all tracks and photons
(excluding the $B$ candidate daughters) flowing into nine concentric
cones centered on the thrust axis of the $B$ candidate, in the CM frame.
Each cone subtends an angle of $10^\circ$ and is folded to combine
the forward and backward intervals.
In addition, the momentum of the highest momentum electron, muon, kaon,
and proton are used as inputs to the Fisher discriminant taking advantage of
the high quality particle identification in CLEO III.

In the case of a candidate mode involving one or more charged pions or
kaons, such as $B\to K\pi$ or $B\to \pi \pi^0$, each charged track must be
positively identified as $K$ or $\pi$. 

\begin{table}[t]
\begin{center}
\begin{tabular}{|lcccc|}
\hline
mode & \multicolumn{4}{c|}{BR ($10^{-6}$) [UL @ 90\% CL]} \\ 
& CLEO & \babar\ & Belle & WA \\ 
\hline 
$B^0 \to \pi^+\pi^-$ 
& $ 4.5 ^{+1.4+0.5}_{-1.2-0.4} $ 
& $ 4.7 \pm 0.6 \pm 0.2 $ 
& $ 4.4 \pm 0.6 \pm 0.3 $ 
& $ 4.6 \pm 0.4  $ \\ 
\hline 
$B^0 \to K^+\pi^-$ 
& $ 18.0 ^{+2.3+1.2}_{-2.1-0.9} $ 
& $ 17.9 \pm 0.9 \pm 0.7 $ 
& $ 18.5 \pm 1.0 \pm 0.7 $ 
& $ 18.2 \pm 0.8  $ \\ 
\hline 
$B^0 \to K^+K^-$ 
& $ < 0.8 $ 
& $ < 0.6 $ 
& $ < 0.7 $ 
& $ < 0.6  $ \\ 
\hline 
$B^+ \to K^0 \pi^+$ 
& $ 18.8 ^{+3.7+2.1}_{-3.3-1.8} $ 
& $ 22.3 \pm 1.7 \pm 1.1 $ 
& $ 22.0 \pm 1.9 \pm 1.1 $ 
& $ 21.8 \pm 1.4  $ \\ 
\hline 
$B^+ \to K^+\pi^0$ 
& $ 12.9 ^{+2.4+1.2}_{-2.2-1.1} $ 
& $ 12.8 ^{+1.2}_{-1.0} \pm 1.0 $ 
& $ 12.8 \pm 1.4 ^{+1.4}_{-1.0} $ 
& $ 12.8 \pm 1.1  $ \\ 
\hline 
$B^0 \to K^0\pi^0$ 
& $ 12.8 ^{+4.0+1.7}_{-3.3-1.4} $ 
& $ 11.4 \pm 1.7 \pm 0.8 $ 
& $ 12.6 \pm 2.4 \pm 1.4 $ 
& $ 11.9 \pm 1.5  $ \\ 
\hline 
$B^+ \to K^0K^+$ 
& $ < 3.3 $ 
& $ < 2.5 $ 
& $ < 3.4 $ 
& $ < 2.5  $ \\ 
\hline 
$B^0 \to K^0\bar K^0$ 
& $ < 3.3 $ 
& $ < 1.8 $ 
& $ < 3.2 $ 
& $ < 1.8  $ \\ 
\hline 
$B^+ \to \pi^+ \pi^0$ 
& $ 4.6 ^{+1.8+0.6}_{-1.6-0.7} $ 
& $ 5.5 ^{+1.0}_{-0.9} \pm 0.6 $ 
& $ 5.3 \pm 1.3 \pm 0.5 $ 
& $ 5.3 \pm 0.8  $ \\ 
\hline 
$B^0 \to \pi^0\pi^0$ 
& $ < 4.4 $ 
& $ < 3.6 $ 
& $ < 4.4 $ 
& $ < 3.6  $ \\ 
\hline
\end{tabular} 
\caption{Branching Fraction results for PP charmless modes: $\pi\pi$, $K\pi$ and $KK$
with all charge combinations~\cite{twobody}. Also the world averages are given.}
\label{table:twobody}
\end{center} 
\end{table}

PID in \babar\ analyses is accomplished with the Cherenkov angle measurement from a
detector of internally reflected Cherenkov light (DIRC).
The final fit to the data includes the normalized Cherenkov residuals 
$\left( \theta_c - \theta_c^\pi\right)/\sigma_{\theta_c}$ and
$\left( \theta_c - \theta_c^K\right)/\sigma_{\theta_c}$, where $\theta_c$ is
the measured Cherenkov angle of the charged primary daughter,
$\sigma_{\theta_c}$ is its error, and
$\theta_c^\pi\,(\theta_c^K)$ is the expected value for a pion(kaon).
The latter two quantities are measured separately for negatively and
positively charged pions and kaons, from a sample of $\Dz\to \Km\pip$ 
originating from $\Dstarp$ decays.

PID in Belle experiment is based on the light yield in the aerogel
Cherenkov counter (ACC) and $dE/dx$ measurements on the central drift chamber.
For each hypothesis ($K$ or $\pi$), the $dE/dx$ and ACC probability density
functions are combined to form likelihoods, ${\cal L}_K$ and ${\cal L}_\pi$.
$K$ or $\pi$ mesons are distinguished by requirements on the likelihood ratio
${\cal L}_K/({\cal L}_K+{\cal L}_\pi)$.

The $K/\pi$ identification in CLEO relies on the pattern of Cherenkov
photon hits in the RICH detector fitted to both a kaon and pion hypothesis,
each with its own likelihood ${\cal L}_K$ and ${\cal L}_\pi$.
Calibrated $dE/dx$ information from the drift chamber is used to compute a
\chisq\ for kaon and pion hypotheses. The RICH and $dE/dx$ results are
combined to form an effective \chisq\ difference, 
$\Delta_{K\pi} = -2\ln{\cal L}_K + 2\ln{\cal L}_\pi + \chisq_K - \chisq_\pi$.
Kaons are identified by $\Delta_{K\pi} < \delta_K$ and pions by $-\Delta_{K\pi} <
\delta_\pi$, with values of $\delta_K$ and $\delta_\pi$ chosen to yield
$(90\pm 3)$\% efficiency as determined in an independent study of tagged
kaons and pions obtained from the decay $D^{*+}\to\pi^+D^0~(D^0\to
K^-\pi^+)$.

Finally PV modes allow for the use of the vector particle helicity angle $\theta_h$
defined as the angle between the direction of one of $h$ daughters
in the $h$ rest frame and the direction of $h$ in the $B$ rest frame.

\section{Results}\label{results}

In general the charmless decay modes taken into account here have contributions
from (a) signal, (b) continuum $q\bar q$ background and (c) cross-feed from
other $B$ modes.

In order to extract the signal yields, \babar\ and CLEO use unbinned extended
maximum likelihood fit. The input variables to the fit are in general $m_{ES}$,
$\Delta E$, Fisher discriminant $\cal F$. Some \babar\ analyses take advantage of a
Neural Network to fight the continuum background instead of the Fisher
discriminant. In CLEO analyses, the flight direction of the $B$ candidate
$\cos\theta_B$ is added to the fit inputs. Moreover most of the charmless
decay analyses in \babar\ include also the Cherenkov angle in the fit to 
distinguish when necessary among differences in the $K/\pi$ content of
the final states.

\begin{table}[t] 
\begin{center} 
\begin{tabular}{|lcccc|}
\hline
mode & \multicolumn{4}{c|}{BR ($10^{-6}$) [UL @ 90\% CL]} \\ 
& CLEO & \babar\ & Belle & WA \\ 
\hline 
$B^0 \to \rho^+\pi^-$ 
& $ 27.6 ^{+8.4}_{-7.4} \pm 4.2 $ 
& $ 22.6 \pm 1.8 \pm 2.2 $ 
& $ 20.8 ^{+6.0+2.8}_{-6.3-3.1} $ 
& $ 22.7 \pm 2.5  $ \\ 
\hline 
$B^+ \to \rho^0\pi^+$ 
& $ 10.4 ^{+3.3}_{-3.4} \pm 2.1 $ 
& $ 24 \pm 8 \pm 3 $ 
& $ 8.0 ^{+2.3}_{-2.0} \pm 0.7 $ 
& $ 9.5 \pm 2.0  $ \\ 
\hline 
$B^+ \to \rho^+\pi^0$ 
& $ < 43 $ 
& $ - $ 
& $ - $ 
& $ < 43  $ \\ 
\hline 
$B^0 \to \rho^0\pi^0$ 
& $ < 5.5 $ 
& $ < 10.6 $ 
& $ < 5.3 $ 
& $ < 5.3  $ \\ 
\hline
$B^0 \to \rho^+K^-$ 
& $ 16 ^{+8}_{-6} \pm 3 $ 
& $ 7.3 ^{+1.3}_{-1.2} \pm 1.2 $ 
& $ < 23 $ 
& $ 7.7 \pm 1.7  $ \\ 
\hline 
$B^+ \to \rho^0K^+$ 
& $ < 17 $ 
& $ < 29 $ 
& $ < 12 $ 
& $ < 12  $ \\ 
\hline 
$B^+ \to \rho^+K^0$ 
& $ < 48 $ 
& $ - $ 
& $ - $ 
& $ < 48  $ \\ 
\hline 
$B^0 \to \rho^0K^0$ 
& $ < 39 $ 
& $ - $ 
& $ < 12 $ 
& $ < 12  $ \\ 
\hline
\end{tabular} 
\caption{Branching Fraction results from the $\rho h$ modes with $h$
being a $\pi$ or a $K$ with all charge combinations.}
\label{table:rho}
\end{center} 
\end{table}

In Belle analyses, the signal yields are extracted by a binned
maximum likelihood fit to the $\Delta E$ distribution in the $m_{ES}$
signal window ($5.271~\gevcc < m_{ES} < 5.289~\gevcc$, the lower limit
being $5.270~\gevcc$ in case of a $\pi^0$ in the final state). The $m_{ES}$
distributions are fitted as a consistency check.

Tables~\ref{table:twobody},~\ref{table:rho} and~\ref{table:phiomega}
show the Branching Fraction results of respectively
the charmless two-body hadronic $B$ decays~\cite{twobody}, the modes with a $\rho$ in the
final states~\cite{rho}, the modes with a $K^*$, a $\phi$ or an $\omega$ in the final
states~\cite{hfag}.
The results shown are from all the three experiments and
also the relative world averages are reported.

\begin{table}[t] 
\begin{center} 
\begin{tabular}{|lcccc|}
\hline
mode & \multicolumn{4}{c|}{BR ($10^{-6}$) [UL @ 90\% CL]} \\ 
& CLEO & \babar\ & Belle & WA \\ 
\hline 
$B^+ \to K^{*0}\pi^+$ 
& $ 7.6 ^{+3.5}_{-3.0} \pm 1.6 $ 
& $ 15.5 \pm 3.4 \pm 1.8 $ 
& $ 19.4 ^{+4.2+4.1}_{-3.9-7.1} $ 
& $ 12.3 \pm 2.6  $ \\ 
\hline 
$B^+ \to K^{*+}\pi^0$ 
& $ < 31 $ 
& $ - $ 
& $ - $ 
& $ < 31  $ \\ 
\hline 
$B^0 \to K^{*+}\pi^-$ 
& $ 16 ^{+6}_{-5} \pm 2 $ 
& $ - $ 
& $ < 30 $ 
& $ 16 \pm 6  $ \\ 
\hline 
$B^0 \to K^{*0}\pi^0$ 
& $ < 3.6 $ 
& $ - $ 
& $ < 7 $ 
& $ < 3.6  $ \\ 
\hline
$B^+ \to \phi K^+$ 
& $ 5.5 ^{+2.1}_{-1.8} \pm 0.6 $ 
& $ 10.0 ^{+0.9}_{-0.8} \pm 0.5 $ 
& $ 9.4 \pm 1.1 \pm 0.7 $ 
& $ 9.3 \pm 0.8  $ \\ 
\hline 
$B^0 \to \phi K^0$ 
& $ 5.4 ^{+3.7}{-2.7} \pm 0.7 $ 
& $ 7.6 ^{+1.3}_{-1.2} \pm 0.5 $ 
& $ 9.0 \pm 2.2 \pm 0.7 $ 
& $ 7.7 \pm 1.1  $ \\ 
\hline 
$B^0 \to \phi K^{*0}$ 
& $ 11.5 ^{+4.5+1.8}_{-3.7-1.7} $ 
& $ 11.1 ^{+1.3}_{-1.2} \pm 0.8 $ 
& $ 10.0 ^{+1.6+0.7}_{-1.5-0.8} $ 
& $ 10.7 \pm 1.1  $ \\ 
\hline 
$B^+ \to \phi K^{*+}$ 
& $ 10.6 ^{+6.4+1.8}_{-4.9-1.6} $ 
& $ 12.1 ^{+2.1}_{-1.9} \pm 1.1 $ 
& $ 6.7 ^{+2.1+0.7}_{-1.9-1.0} $ 
& $ 9.4 \pm 1.6  $ \\ 
\hline 
$B^+ \to \omega K^+$ 
& $ 3.2 ^{+2.4}_{-1.9} \pm 0.8 $ 
& $ 5.0 \pm 1.0 \pm 0.4 $ 
& $ 6.7 ^{+1.3}_{-1.2} \pm 0.6 $ 
& $ 5.4 \pm 0.8  $ \\ 
\hline 
$B^+ \to \omega \pi^+$ 
& $ 1.3 ^{+3.3}_{-2.9} \pm 1.4 $ 
& $ 5.4 \pm 1.0 \pm 0.5 $ 
& $ 5.9 ^{+1.4}_{-1.3} \pm 0.6 $ 
& $ 5.3 \pm 0.9  $ \\ 
\hline 
$B^0 \to \omega K^0$ 
& $ 10.0 ^{+5.4}_{-4.2} \pm 1.4 $ 
& $ 5.3 ^{+1.4}_{-1.2} \pm 0.5 $ 
& $ < 7.6 $ 
& $ 5.6 \pm 1.4  $ \\ 
\hline
\end{tabular} 
\caption{Branching Fraction results for PV final states including $K^*$, $\phi$ or $\omega$.}
\label{table:phiomega}
\end{center} 
\end{table}

In Figure~\ref{fig:twobody} plots are shown from the various two-body analyses
and techniques. The top right left are two examples from \babar\ analyses and represent
the projections of the $m_{ES}$ distributions from the unbinned maximum likelihood
fit for events that satisfy optimized requirements on probability ratios
for signal to background based on all variables except $m_{ES}$ itself.
The bottom left plots show the distributions of
$-2\ln({\cal L}/{\cal L}_{\rm max})$ for CLEO II and CLEO III combined
for the $K\pi$ and $\pi\pi$ modes with non-zero yields.
The remaining right plots are two examples from Belle and show the $\Delta E$ distributions
in the $m_{ES}$ signal region for $\pi^+\pi^-$ and $K^+\pi^-$ respectively.
The results of the fits used to extract the signal yields are also shown.

In Table~\ref{table:asymmetry} the asymmetry results are shown from all three
collaborations. All the results are compatible with zero and still
statistical dominated.

The additional statistics that will be collected
in the next years by \babar\ and Belle will provide exciting results in the
physics of charmless decay.

\begin{table}[tbhp] 
\begin{center}
\begin{tabular}{|lcccc|}
\hline
mode & CLEO & \babar\ & Belle & WA \\ 
\hline 
$B^0 \to K^+ \pi^-$ 
& $ -0.04 \pm 0.16 \pm 0.02 $ 
& $ -0.10 \pm 0.04 \pm 0.01 $ 
& $ -0.07 \pm 0.06 \pm 0.01 $ 
& $ -0.09 \pm 0.03 $ \\ 
\hline 
$B^+ \to K^+ \pi^0$ 
& $ -0.29 \pm 0.23 \pm 0.02 $ 
& $ -0.09 \pm 0.09 \pm 0.01 $ 
& $ ~~0.23 \pm 0.11 ^{+0.01}_{-0.04} $ 
& $ ~~0.00 \pm 0.07 $ \\ 
\hline 
$B^0 \to K^0 \pi^+$ 
& $ ~~0.18 \pm 0.24 \pm 0.02 $ 
& $ -0.05 \pm 0.08 \pm 0.01 $ 
& $ ~~0.07^{+0.09}_{-0.08}$$^{+0.01}_{-0.03} $ 
& $ ~~0.01 \pm 0.06 $ \\ 
\hline 
$B^+ \to K^0 \pi^0$ 
& $ - $ 
& $ ~~0.03 \pm 0.36 \pm 0.09 $ 
& $ - $ 
& $ ~~0.03 \pm 0.37 $ \\ 
\hline 
$B^+ \to \pi^+ \pi^-$ 
& $ - $ 
& $ ~~0.30 \pm 0.25 \pm 0.04 $ 
& $ ~~0.77 \pm 0.27 \pm 0.08 $ 
& $ ~~0.51 \pm 0.19 $ \\ 
\hline
$B^+ \to \pi^+ \pi^0$ 
& $ - $ 
& $ -0.03 ^{+0.18}_{-0.17} \pm 0.02 $ 
& $ -0.14 \pm 0.24 ^{+0.05}_{-0.04} $ 
& $ -0.07 \pm 0.15 $ \\ 
\hline
$B^+ \to \rho^-\pi^+$ 
& $ - $ 
& $ -0.11 ^{+0.16}_{-0.17} \pm 0.04 $ 
& $ - $ 
& $ -0.11 \pm 0.17 $ \\ 
\hline
$B^+ \to \rho^+\pi^-$ 
& $ - $ 
& $ -0.62 ^{+0.24}_{-0.28} \pm 0.06 $ 
& $ - $ 
& $ -0.62 \pm 0.29 $ \\ 
\hline
$B^+ \to K^+ \rho^-$ 
& $ - $ 
& $ ~~0.28 \pm 0.17 \pm 0.08 $ 
& $ - $ 
& $ ~~0.28 \pm 0.19 $ \\ 
\hline
\end{tabular} 
\caption{Direct \CP\ asymmetry results for $K\pi$, $\pi\pi$, $\rho\pi$ and $\rho K$ modes}
\label{table:asymmetry}
\end{center} 
\end{table}

\label{BonaEnd}
 
\end{document}